\documentclass{llncs}
\usepackage{amsfonts}
\usepackage{amsmath}
\usepackage{latexsym}
\input epsf.sty
\usepackage{url}
\usepackage{theorem}

\newcommand{\ket}[1]{|#1\rangle}

\def\01{\{0, 1\}}

\newcommand{\comment}[1]{}

\begin{document}

\newcommand{\tinyspace}{\mskip 1mu}
\title{New Developments in Quantum Algorithms}
\author{%
Andris Ambainis
  \thanks{Supported by FP7 Marie Curie Grant PIRG02-GA-2007-224886 and ESF project 1DP/1.1.1.2.0/09/APIA/VIAA/044.}
}
\authorrunning{Andris Ambainis}
\tocauthor{Andris Ambainis (University of Latvia)}
\institute{%
Faculty of Computing,
University of Latvia, Raina bulv. 19, Riga, LV-1586, Latvia,
  \email{andris.ambainis@lu.lv}
}
\date{}
\maketitle

\begin{abstract}
In this talk, we describe two recent developments in quantum algorithms.

The first new development is a quantum algorithm for evaluating a Boolean formula consisting
of AND and OR gates of size $N$ in time $O(\sqrt{N})$. 
This provides quantum speedups for any problem that can be 
expressed via Boolean formulas. This result can be also extended to 
{\em span problems}, a generalization of Boolean formulas.
This provides an optimal quantum algorithm for any Boolean function in the
black-box query model.

The second new development is a quantum algorithm for solving systems of linear 
equations. In contrast with traditional algorithms that run in time $O(N^{2.37...})$
where $N$ is the size of the system, the quantum algorithm runs in time $O(\log^c N)$.
It outputs a quantum state describing the solution of the system.
\end{abstract}

\section{History of quantum algorithms}

\subsection{First quantum algorithms}

Quantum computing (and, more broadly, quantum information science) is a new area at the boundary of computer science and physics. It studies how to apply quantum mechanics to solve problems in computer science and information processing.
The area of quantum computing was shaped by the discoveries of two major 
quantum algorithms in mid-1990s. 

The first of the these two discoveries was Shor's 
polynomial time quantum algorithm for factoring and discrete logarithms.
Factoring and discrete logarithm are very hard number theoretic problems. 
The difficulty of these problems has been used to design cryptosystems 
(such as RSA and Diffie-Helman key exchange) for secure data transmission 
over an insecure network (such as Internet). The security of data 
transmission is based on the assumption that it is hard to factor (or find 
discrete logarithm of) large numbers. Until recently, this assumption was 
not in doubt. Mathematicians had tried to devise an efficient way of 
factoring large numbers for centuries, with no success.

In 1994, Shor \cite{Shor} discovered a fast algorithm for factoring large numbers - on a quantum mechanical computer. This shook up the foundations of cryptography. If a quantum mechanical computer is built, today's methods for secure data transmission over the Internet will become insecure.

Another, equally strikingly discovery was made in 1996, by Lov Grover \cite{Grover}. He invented a quantum algorithm for speeding up exhaustive search problems. Grover's algorithm solves a
generic exhaustive search problem with $N$ possible solutions in time $O(\sqrt{N})$.
This provides a quadratic speedup for a range of search problems, from problems
that are in P classically to NP-complete problems.

Since then, each of the two algorithms has been analyzed in great detail.
Shor's algorithm has been generalized to solve a class of algebraic problems
that can be abstracted to {\em Abelian hidden subgroup problem} \cite{Jozsa}. 
Besides factoring
and discrete logarithm, the instances of Abelian HSP include 
cryptanalysis of hidden linear equations \cite{BL}, solving 
Pell's equation, 
principal ideal problem \cite{Hallgren} and others.

Grover's algorithm has been generalized to the framework of 
{\em amplitude amplification} \cite{BHMT} and extended to solve problems like
approximate counting \cite{BHT,NW} and collision-finding \cite{BHT1}.

\subsection{Quantum walks and adiabatic algorithms}

Later, two new methods for designing quantum algorithms emerged:
quantum walks \cite{Amb04surv,Kempe,Amb08surv,Santha,Venegas} and adiabatic algorithms 
\cite{Farhi}.

Quantum walks are quantum generalizations of classical random walks. 
They have been used to obtain quantum speedups for a number of problems.
The typical setting is as follows. 
Assume that we have a classical Markov chain, on a state-space 
in which some states are special (marked). The Markov chain
starts in a uniformly random state and stops if it reaches a marked state.
If the classical Markov chain reaches a marked state in expected time $T$,
then there is a quantum algorithm which can find it in time $O(\sqrt{T})$,
assuming some conditions on the Markov chain
\cite{Szegedy,MNRS,KMOR}.

This approach gives quantum speedups for a number of problems: element distinctness
\cite{Ambainis04}, search on a grid \cite{AKR,Tulsi}, finding triangles in graphs \cite{MSS},
testing matrix multiplication \cite{BS06} and others.

Another application of quantum walks is to the "glued trees" problem \cite{CC+}.
In this problem, we have a graph $G$ with two particular vertices
$u, v$, designed as the entrance and the exit. The problem is
to find the vertex $v$, if we start at the vertex $u$. 
There is a special exponential size graph called "glued trees" 
on which any classical algorithm needs exponential time to find $v$
but a quantum algorithm can find $v$ in polynomial time \cite{CC+}.

Adiabatic computation is a physics-based paradigm for quantum algorithms.
In this paradigm, we design two quantum systems:
\begin{itemize}
\item
$H_{sol}$ whose lowest-energy state $\ket{\psi_{sol}}$ encodes a solution 
to a computational problem (for example, a satisfying assignment
for SAT). 
\item
$H_{start}$ whose lowest-energy state $\ket{\psi_{start}}$ is such that we can easily
prepare $\ket{\psi_{start}}$.
\end{itemize}
We then prepare $\ket{\psi_{start}}$ and slowly transform the forces 
acting on the quantum system from $H_{start}$ to $H_{sol}$.
Adiabatic theorem of quantum mechanics guarantees that, if the transformation
is slow enough, $\ket{\psi_{start}}$ is transformed into a state
close to $\ket{\psi_{sol}}$ \cite{Farhi}.

The key question here is: what is "slowly enough"? 
Do we need a polynomial time or an exponential time to transform
$H_{start}$ to $H_{sol}$ (thus solving SAT by a quantum algorithm)?
This is a subject of an ongoing debate \cite{Farhi,DMV,AKR09}.

Adiabatic computation has been used by D-Wave Systems \cite{DW} which claims
to have built a 128-bit adiabatic quantum computer. 
However, the claims of D-Wave have been questioned by 
many prominent scientists (see e.g. \cite{Aar}).

\subsection{Most recent algorithms}

Two most recent discoveries in this field are the quantum algorithms for
formula evaluation \cite{FGG} and solving systems of linear equations \cite{HHL}.
Both of those algorithms use the methods from the previous algorithms but do it in
a novel, unexpected way. Formula evaluation uses quantum walks but in a form
that is quite different from the previous approach (which we described above). 
Quantum algorithm for formula evaluation uses {\em eigenvalue estimation} \cite{ME}
which is the key technical subroutine of Shor's factoring algorithm \cite{Shor}
and the related quantum algorithms. 
But, again, eigenvalue estimation is used in a very unexpected way.

These two algorithms are the main focus of this survey. We describe them in detail
in sections \ref{sec:formula} and \ref{sec:equations}.

\section{Formula evaluation}
\label{sec:formula}

\subsection{Overview}

We consider evaluating a Boolean formula of variables $x_1, \ldots, x_N$
consisting of ANDs and ORs, with each variable occuring exactly once in the formula.
Such a formula can be described by a tree, with variables $x_i$ at the leaves 
and AND/OR gates at the internal nodes. 
This problem has many applications because Boolean formulas can be used to
describe a number of different situations. The most obvious one is 
determining if the input data $x_1, \ldots, x_N$ satisfy certain
constraints that can be expressed by AND/OR gates.

For a less obvious application, we can view
formula evaluation as a black-box model for 
a 2-player game (such as chess) if both players play their optimal strategies. 
In this case, the game can be represented by a game tree consisting
of possible positions. The leaves of a tree correspond to the possible end
positions of the game. Each of them contains a variable $x_i$, with $x_i=1$
if the $1^{\rm st}$ player wins and $x_i=0$ otherwise. If an internal node
$v$ corresponds to a position in which the $1^{\rm st}$ player makes the next move,
then $v$ contains a value that is OR of the values of $v$'s children. 
(The $1^{\rm st}$ player wins if he has a move 
that leads to a position from which he can win.) If $v$ is a node for 
which the $2^{\rm nd}$ player makes the next move, then $v$ contains a value
that is AND of the values of $v$'s children. (In this case, 
the $1^{\rm st}$ player wins if he wins
for any possible move of the $2^{\rm nd}$ player.)

The question is: assuming we have no further information about the game 
beyond the position tree, how many of the variables $x_i$ do we have
to examine to determine whether the $1^{\rm st}$ player has a winning strategy?

Classically, the most widely studied case is the full binary tree of
depth $d$, with $N=2^d$ leaves. It can be evaluated by looking
at $\Theta(N^{.754...})$ leaves and this is optimal \cite{Santha,SW,Snir}.
A natural question was whether one could achieve a better result,
using quantum algorithms. This question was a well known open problem 
in the quantum computing community since mid-1990s. Until 2007, the
only known result was that $\Omega(\sqrt{N})$ quantum steps are necessary,
for any AND-OR tree \cite{Amb00,BS}.

\subsection{The model}

By standard rules from Boolean logic (de Morgan's laws), we can replace both AND 
and OR gates by NAND gates. A NAND gate $NAND(y_1, \ldots, y_k)$ outputs 1 
if $AND(y_1, \ldots, y_k)=0$ (i.e., $y_i=0$ for at least one $i\in\{1, \ldots, k\}$)
and 0 otherwise. Then, we have a tree with $x_1, \ldots, x_N$ at the leaves
and NAND gates at the internal vertices. The advantage of this transformation
is that we now have to deal with just one type of logic gates (instead of two - AND 
and OR).

We work in the quantum query model. In the discrete-time version 
of this model \cite{Ambainis-survey,BWSurvey}, the input bits $x_1, \ldots, x_N$ 
can be accessed by queries $O$ to a  black box.

To define $O$, we represent basis states as $|i, z\rangle$ where
$i\in\{0, 1, \ldots, N\}$. The query transformation $O_x$ 
(where $x=(x_1, \ldots, x_N)$) maps $\ket{0, z}$ to $\ket{0, z}$ and 
$\ket{i, z}$ to $(-1)^{x_i}\ket{i, z}$ for $i\in\{1, ..., N\}$
(i.e., we change phase depending on $x_i$, unless $i=0$ in which case we do
nothing).

Our algorithm can involve queries $O_x$ and arbitrary non-query transformations
that do not depend on $x_1, \ldots, x_N$. The task is to solve a 
computational problem (e.g., to compute a value of a NAND formula)
with as few queries as possible.

\subsection{Results}

In 2007, in a breakthrough result, Farhi et al. \cite{FGG} showed that  
the full binary AND-OR tree can be evaluated in $O(\sqrt{N})$ quantum time
in continuous-time counterpart of the query model.

Several improvements followed soon.
Ambainis et al. \cite{Childs-v1,Ambainis-NAND,AC+} translated the algorithm of
\cite{FGG} to the conventional discrete time quantum query model and
extended it to evaluating arbitrary Boolean formulas with $O(N^{1/2+o(1)})$ 
quantum queries. 

Soon after, Reichardt and \v Spalek \cite{RS} discovered a far-reaching 
generalization of this result. Namely, the quantum algorithm was generalized 
to evaluating span programs. 
A span program is an algebraic model of computation, originally 
invented for proving lower bounds on circuit size \cite{KW93}.

In a span program, we have a {\em target} vector $v$ in some linear space.
We also have other vectors $v_1, \ldots, v_m$, each of which is 
associated with some condition $x_i=0$ or $x_i=1$. 
The span program evaluates to 1 on an input $x_1, \ldots, x_n$ if
$v$ is equal to a linear combination of vectors 
$v_{i_1}, \ldots, v_{i_k}$ which are associated with conditions 
that are true for the given input $x_1, \ldots, x_n$.
Otherwise, the span program evaluates to 0. 

Here is an example of a span program. We have a two dimensional linear space,
with the following vectors:
\[ v = \left( \begin{array}{c} 1 \\ 0 \end{array} \right), 
v_1 = \left( \begin{array}{c} 1 \\ a \end{array} \right),
v_2 = \left( \begin{array}{c} 1 \\ b \end{array} \right),
v_3 = \left( \begin{array}{c} 1 \\ c \end{array} \right) \]
where $a, b, c$ are any three distinct non-zero numbers.
Vectors $v_1, v_2, v_3$ are associated with conditions 
$x_1=1$, $x_2=1$, $x_3=1$, respectively.

Given any two of $v_1, v_2, v_3$, we can express any vector in two dimensions 
(including $v$) as their linear combination. Thus, this span program 
computes the majority function $MAJ(x_1, x_2, x_3)$ which is 1 whenever
at least 2 of variables $x_1, x_2, x_3$ are equal to 1.

Logic formulae can be embedded into span programs.
That is, if we have two span programs computing functions $f_1(x_1, \ldots, x_N)$
and $f_2(x_1, \ldots, x_N)$, we can combine them into span programs for
$f_1$ AND $f_2$ and $f_1$ OR $f_2$ in a fairly simple way.
\comment{For example, the span program for $f_1 AND f_2$ 
is obtained as follows. Let $d_1$ and $d_2$ be the dimensions 
of span programs for $f_1$ and $f_2$. We then create a span program in $d=d_1+d_2+1$ 
dimensions in a following way:
\begin{itemize}
\item
We take each vector $v_i$ from the span program for $f_1(x_1, \ldots, x_N)$ and extend it to
a vector in $d$ dimensions by adding 0s in the last $d_2+1$ coordinates.
\item
We take each vector $v_i$ from the span program for $f_2(x_1, \ldots, x_N)$ and extend it to
a vector in $d$ dimensions by adding 0s in the first $d_1$ coordinates and the last coordinate.
\item
We add a vector $v'$ with 1}

Reichardt and \v Spalek \cite{RS} invented a complexity measure, {\em witness size} for span
programs. This measure generalizes formula size: 
a logic formula of size $S$ can be transformed into a span program with
witness size $S$. \cite{RS} gave a quantum algorithm for
evaluating a span program of witness size $S$ with $O(\sqrt{S})$ queries.
This is a very interesting result because it allows to evaluate formulas with gates
other than AND and OR by designing span programs for those gates and them composing them into one 
big span program. The next step was even more interesting.

The next step was even more surprising. Reichardt \cite{R2009,R2009a,R2010} discovered that the span
program approach is optimal, for any Boolean function $f(x_1, \ldots, x_N)$. 
That is \cite{R2010}, if $Q(f)$ is the minimum number of quantum queries for evaluating
$f$ (by any quantum algorithm), then there is a span program with witness size $O(Q^2(f))$.
Thus, a span-program based algorithm can evaluate $f$ with $O(Q(f))$ queries, within
a constant factor of the best possible number of queries.

This fact linked two lines of research: quantum formula evaluation algorithms and "quantum adversary"
lower bounds. "Quantum adversary" (invented in \cite{Amb00}) is a method for proving 
lower bounds on the number of queries to evaluate $f(x_1, \ldots, x_N)$ by a quantum algorithm.
Several progressively stronger versions of "quantum adversary" have been invented 
\cite{Ambainis-poly,BSS,LM,HLS}, with the strongest being the "negative adversary" method
of \cite{HLS}. 

Finding the best lower bound for quantum algorithms provable via "negative adversary" method is
a semidefinite program (SDP). Reichardt \cite{R2009,R2009a,R2010} considered the dual of this SDP and
showed that the dual SDP gives the span program with the smallest witness size. 
Thus, the span programs are optimal (in terms of query complexity) for any Boolean function
$f(x_1, \ldots, x_N)$. (Finding the optimal span program, however, requires solving a semidefinite
program of size $2^N$.)

As a by-product, this gave a better algorithm for formula evaluation, improving the complexity from
$O(N^{1/2+o(1)})$ in \cite{AC+} to $O(\sqrt{N} \log N)$ in \cite{R2009b}.

\subsection{Algorithmic ideas}

We now give a very informal sketch the simplest version of formula evaluation algorithm.
We augment the formula tree with a finite "tail" of length $L$ as shown in Figure \ref{myfig}.
\begin{figure*}
\begin{center}
\epsfxsize=3in
\hspace{0in}
\epsfbox{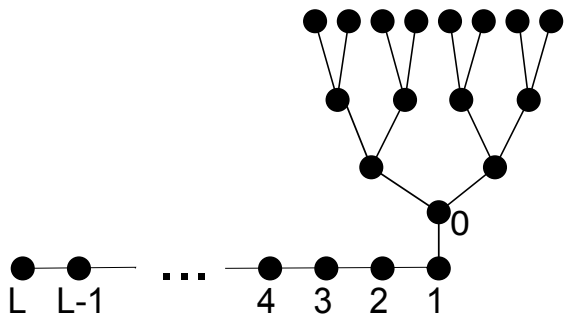}
\caption{A formula tree augmented with a finite "tail"}
\label{myfig}
\end{center}
\end{figure*}
We then consider a quantum walk on this tree. At the leaves of the tree, 
the transformations that are performed depend on whether the leaf holds $x_i=0$
or $x_i=1$. (This is achieved by querying the respective $x_i$ and then performing 
one of two transformations, depending on the outcome of the query.) 

The starting state is an appropriately chosen quantum state 
$\ket{\psi_{start}}=\sum_i \alpha_i \ket{i}$
consisting of the states $\ket{i}$ in the tail. If the quantum walk is set up properly,
an amazing thing happens! Whenever the formula evaluates to 0, the state $\ket{\psi_{start}}$
remains almost unchanged. Whenever the formula evaluates to 1, after $O(N^{1/2+o(1)})$ steps, the 
state is almost completely different from $\ket{\psi_{start}}$. This means that we can distinguish
between the two cases by running the walk for $O(N^{1/2+o(1)})$ steps and measuring whether the state
is still $\ket{\psi_{start}}$. Surprisingly, the behaviour of the walk only depends
on the value of the formula and not on which particular variables $x_1, \ldots, x_N$ are 1.

The algorithm for evaluating span programs is essentially the same, except that the quantum walk
is performed on a weighted graph that corresponds to the span program.

For more information on this topic, we refer the reader to the survey \cite{Amb09-surv}
and the original papers.

\section{Linear equations}
\label{sec:equations}

\subsection{Overview}

Solving large systems of linear equations is a very common problem
in scientific computing, with many applications.
We consider solving a system of $N$ linear equations with $N$ unknowns: $Ax=b$ where
\[ 
A= \left( \begin{array}{cccc} 
a_{11} & a_{12} & \ldots & a_{1N} \\
a_{21} & a_{22} & \ldots & a_{2N} \\
\ldots & \ldots & \ldots & \ldots \\
a_{N1} & a_{N2} & \ldots & a_{NN} 
\end{array}
\right), 
x= \left( \begin{array}{c} 
x_1 \\
x_2 \\
\ldots \\
x_N
\end{array}
\right),
b= \left( \begin{array}{c} 
b_1 \\
b_2 \\
\ldots \\
b_N
\end{array}
\right) .\]
$A$ and $b$ are given to us. The task is to find $x$.

The best classical algorithm for solving a general system  
$Ax=b$ runs in time $O(N^{2.37...})$.
The reason for that was that even 
outputting the solution requires time $\Omega(N)$
because the solution contains values for $N$ variables.
Thus, it seemed that there was no hope for achieving 
more than a polynomial speedup by a quantum algorithm.
 
Recently, Harrow, Hassidim and Lloyd \cite{HHL} discovered a 
surprising quantum algorithm that allows to solve systems of
linear equations in time $O(\log^c N)$ - in an unconventional
sense. Namely, the algorithm of \cite{HHL} generates the quantum state
\[ \ket{\psi}=\sum_{i=1}^N x_i \ket{i} \] 
with the coefficients $x_i$ being
equal to the values of variables in the solution $x=(x_1, x_2, \ldots, x_N)$
of the system $Ax=b$.

What can we do with this quantum state? We cannot extract all the values $x_i$ from it.
If we measured this state, we would obtain one value $i$, with probabilities of 
different $i$ proportional to $|x_i|^2$.

We can, however, estimate some quantities that depend on all of $x_i$.
For example, if all variables in the solution had values 1 or -1, having the quantum
state $\ket{\psi}$ would enable us to estimate the fraction of variables $x_i=-1$.
Moreover, similar tasks appear to be hard classically. As shown by \cite{HHL},
a classical $O(\log^c N)$-time algorithm for computing any quantity of this type 
implies a polynomial time classical algorithm for simulating any quantum computation.
Thus (unless P=BQP), this quantum algorithm provides a genuine speedup over the classical
algorithms.

\subsection{More details}

In more detail, the running times of both classical and quantum algorithms 
for solving systems of linear equations actually depend on several parameters.
One parameter is $N$, the number of equations (and variables). 
Another parameter is
$\kappa$, the condition number of the system. $\kappa$ is defined as 
$\frac{\mu_{max}}{\mu_{min}}$ where $\mu_{min}$ and $\mu_{max}$ are the 
smallest and the largest singular values of the matrix $A$ 
\cite[Chapter 5.8]{HJ}.

Intuitively, the condition number describes the closeness of the matrix $A$ 
to a singular matrix. For a singular matrix, $\mu_{min}=0$ and $\kappa=\infty$.
Larger condition number means that the matrix $A$ is closer to a singular matrix.
In this case, small changes to input data $A$ and $b$ (or small numerical inaccuracies)
can cause large changes to solution $x$. To compensate, if we have a matrix $A$
with large $\kappa$, we have to perform the computation with a higher accuracy.
This increases the running time of all algorithms for solving systems of linear
equations but the exact increase varies greatly from algorithm to algorithm.

The main classical algorithms for solving systems of linear equations are:
\begin{enumerate}
\item
LU-decomposition \cite[Chapter 3]{GL} which runs in time $O(N^{2.376...} \log^c(\kappa))$ \cite{BH}.
Here, $w=2.376...$ is the constant from the running time $O(N^w)$ of the best matrix multiplication
algorithm \cite{CW}.
\item
Conjugate gradient \cite[Chapter 10]{GL}, which runs in time $O(m \sqrt{\kappa})$ \cite{S94}
where $m$ is the number of non-zero entries in the matrix. If we know that each row of $A$
contains at most $s$ non-zero entries, this is at most $O(N s \sqrt{\kappa})$.
\end{enumerate}

The running time of the quantum algorithm \cite{HHL} is $O(\kappa^2 T \log^c N )$ 
where $T$ is the time necessary to implement the transformation $e^{iA}$ on a quantum computer.
$T$ varies greatly, depending on $A$. For sparse $A$ with at most
$s$ nonzero values in each row and each column, $T=O(s^2 \log N)$ \cite{Ahokas}.
Thus, in this case the running time of the quantum algorithm is $O(\kappa^2 s^4 \log^c N)$.
This achieves an exponential speedup for the case when $N$ is large and $\kappa$ is relatively
small (e.g., $\kappa=O(1)$ or $\kappa=O(\log N)$).

The key bottleneck is the dependence on $\kappa$ which is actually worse than in the classical
algorithms. We have been able to improve it to
$O(\kappa^{1+o(1)} \log^c N)$ \cite{Amb10}.
Unfortunately, further improvement is very unlikely.
\cite{HHL} have shown that an $O(\kappa^{1-\epsilon} \log^c N)$ time quantum algorithm
would imply BQP=PSPACE.

For non-sparse $A$, one could use the algorithms of \cite{Childs,BC} to simulate $e^{iA}$.
The dependence on $N$ is better than $O(N^{2.376...})$ in the classical LU decomposition
but the speedup is only polynomial.

\end{document}